\documentclass[conference,a4paper]{IEEEtran}

\IEEEoverridecommandlockouts

\ifCLASSINFOpdf

\else

\fi

    \newcommand{\eref}[1]{(\ref{#1})}               
    \newcommand{\fig}[1]{Fig. \ref{#1}}           
    
    \newtheorem{remark}{Remark}
    \newtheorem{example}{Example}
    
    \newcommand{\given}{\:\!\vert\:\!}
    \newcommand*{\defeq}{\mathrel{\rlap{%
    			\raisebox{0.3ex}{$\m@th\cdot$}}%
    		\raisebox{-0.3ex}{$\m@th\cdot$}}%
    	=}
\usepackage[english]{babel}
\usepackage[latin1]{inputenc}
\usepackage{times}
\usepackage{csquotes}
\usepackage{pdfpages}
\usepackage{siunitx}
\usepackage[T1]{fontenc}
\usepackage{verbatim,adjustbox}
\usepackage{tikz}
\usetikzlibrary{arrows.meta}
\usepackage{verbatim}
\usepackage{amssymb}
\usetikzlibrary{calc}
\usepackage{tikzscale}
\usepackage{pgfplots}
\usepackage{graphicx,cite,amssymb}
\usepackage[tbtags]{amsmath}
\usepackage{tikz}
\usepackage{diagbox}
\usepackage{lipsum,adjustbox}
\usepackage[nolist]{acronym}
	\tikzstyle{block}=[rectangle, draw, minimum height=4em, minimum width=9em]
	\tikzstyle{init} = [pin edge={to-,thin,black}]
\usepackage{stackengine}

\addto\captionsenglish{}
\begin{document}
\title{Polar-Coded Pulse Position Modulation\\ for the Poisson Channel}
\author{\IEEEauthorblockN{Delcho Donev}
\IEEEauthorblockA{Institute for Communications Engineering\\
Technical University of Munich\\
Email: delcho.donev@tum.de}
\and
\IEEEauthorblockN{Georg B{\"o}cherer}
\IEEEauthorblockA{Mathematical and Algorithmic Sciences Lab\\
	Huawei Technologies France\\
	Email: georg.boecherer@ieee.org}}

\maketitle

\begin{abstract}
A polar-coded modulation scheme for deep-space optical communication is proposed. The photon counting Poisson channel with pulse position modulation (PPM) is considered. We use the fact that PPM is particularly well suited to be used with multilevel codes to design a polar-coded modulation scheme for the system in consideration. The construction of polar codes for the Poisson channel based on Gaussian approximation is demonstrated to be accurate. The proposed scheme uses a cyclic redundancy check outer code and a successive cancellation decoder with list decoding and it is shown that it outperforms the competing schemes.
\end{abstract}

\IEEEpeerreviewmaketitle

\section{Introduction}
\label{sec:Introduction}
This paper designs polar codes for the pulse position modulation (PPM) Poisson channel. This channel models a deep-space, direct-detection optical communications link with a photon counting detector \cite{Barsoum10}. It has been shown in \cite{Hemmati05} that for the deep-space Poisson channel, when operating at low signal to noise ratio (SNR), PPM is near optimal. The combination of a binary error-correcting code with a higher-order modulation scheme can be done in several ways, e.g., with bit-interleaved coded modulation (BICM) \cite{Zehavi, Caire98} or multi-level coding (MLC) with multi-stage decoding (MLC-MD) \cite{Imai}. BICM uses bit-metric decoding (BMD), which calculates bit-wise log-likelihood ratios (LLR), which are then processed independently. However, the LLRs calculated from the same channel output are dependent \cite[Sec. II-D]{Georg16}. This means that BICM cannot always approach the coded modulation (CM) capacity.

Following \cite{Barsoum10}, we plot in ~\fig{fig:cm_vs_bicm1} the PPM capacity \eref{eq:capacity} and the achievable BMD rate (see \eref{eqn:bmrate} below) as a function of the average received power per slot $P_\text{av}=\frac{n_\text{s}}{M}$ in \si{dB} for $n_{\text{b}}=0.2$ where $n_{\text{b}}$ is the average number of noise photons at the receiver (the channel model is explained in Sec. II). Note that, for a wide range of $P_\text{av}$, the BMD rate is significantly lower than the PPM capacity. This has motivated previous works to introduce BICM with iterative demapping (BICM-ID), between the decoder and the PPM detector \cite{Barsoum10}. In \cite{Barsoum10} both convolutional codes (CC) and low-density parity-check (LDPC) codes were studied for the PPM Poisson channel. Another approach is to use non-binary error correction codes in combination with PPM \cite{Balazs17}.

In this paper, we use the fact that PPM modulation calls for multilevel codes (MLC) and this is naturally provided by polar-coded modulation (PCM) \cite{Seidl13, Georg16}. We use polar codes with a successive cancellation (SC) decoder with list decoding (LS) and a cyclic redundancy check (CRC) outer codes \cite{Tal15}. We use the Gaussian approximation construction to design polar codes for higher-order modulation \cite{Georg16}. We show that our proposed scheme improves the bit-error rate (BER) as compared to the scheme proposed in \cite{Barsoum10} for the block length $8208$ bits, CRC size of $14$ and a dynamic list size with at most $16384$ entries. An operating point of our scheme is shown in ~\fig{fig:cm_vs_bicm1}, which shows that we are operating close to the PPM capacity at BER $=3.2\times 10^{-3}$.
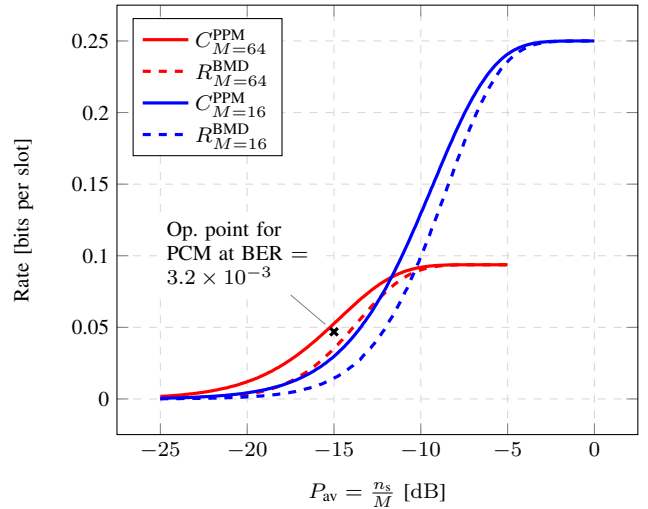
\begin{figure}[!t]
	\centering
	\footnotesize
	\begin{tikzpicture}[]
	\begin{axis}[xlabel={$P_\text{av}=\frac{n_\text{s}}{M}$ [\si{dB}]}, ylabel={Rate [bits per slot]},legend cell align=left,legend entries={$C^{\text{PPM}}_{M=64}$,$R^{\text{BMD}}_{M=64}$,
		$C^{\text{PPM}}_{M=16}$,$R^{\text{BMD}}_{M=16}$
		},legend pos= north west,grid = both,grid style={dashed, gray!30},ytick={0,0.05,0.1,0.15,0.2,0.25},yticklabels={0,0.05,0.1,0.15,0.2,0.25},every axis plot/.append style={very thick}]
	\addplot[red]  table [x={P}, y={cm}] {cmvsbicm64.dat};
	\addplot[red, dashed]  table [x={P}, y={bicm}] {cmvsbicm64.dat};
	\addplot[blue]  table [x={P}, y={cm}] {cmvsbicm16.dat};
	\addplot[blue, dashed]  table [x={P}, y={bicm}] {cmvsbicm16.dat};
		\addplot[mark=x] coordinates {(-15,0.0469)} node[pin={[text width=2cm,fill=white]100:{Op. point for PCM at $\text{BER}=3.2\times 10^{-3}$}}]{};
	\end{axis}
	\end{tikzpicture}
	\caption{PPM capacity \eref{eq:cap} and BMD rate \eref{eqn:bmrate} for $n_{\text{b}}=0.2$. An operating point for the PCM scheme at $P_{\text{av}}=$\SI{-15}{dB} and BER $=3.2\times 10^{-3}$ is marked.}
	\label{fig:cm_vs_bicm1}
\end{figure}


This paper is organized as follows. PCM for the PPM Poisson channel is explained in Section~II. Section~III describes the polar code construction method for the PPM Poisson channel. In Section~IV the experimental results are presented and Section~V concludes the paper.

\section{PPM Poisson Channel}

\subsection{Channel Model}
\label{sec:ChannelModel}

\begin{figure*}[!t]
	\centering
	\begin{adjustbox}{width=\textwidth}
		\begin{tikzpicture}[auto]
		\node [block] (Modulator) {$M$-PPM Modulator};
		\node [block] (Encoder) [left of=Modulator, node distance=20em] {Encoder};
		\node (Input) [left of=Encoder, node distance=16em] {};
		\node [block] (Channel) [right of=Modulator, node distance=22em] {Poisson Channel};
		\node (Output) [right of=Channel, node distance=16em] {};
		\path[-{Latex[length=3mm]}] (Encoder) edge node [above] {$\tilde{\boldsymbol{C}}$} (Modulator);
		\path[-{Latex[length=3mm]}] (Input) edge node [above] {$\tilde{\boldsymbol{U}}$} (Encoder);
		\path[-{Latex[length=3mm]}] (Input) edge node [below] {\begin{tabular}{c}$\boldsymbol{U}_1,\boldsymbol{U}_2,\dotsc,\boldsymbol{U}_{m}$ \\ $mn$-bit vector\end{tabular}} (Encoder);
		\path[-{Latex[length=3mm]}] (Encoder) edge node [below] {\begin{tabular}{c}$\boldsymbol{C}_1,\boldsymbol{C}_2,\dotsc,\boldsymbol{C}_{m}$ \\ $mn$-bit vector\end{tabular}} (Modulator);
		\path[-{Latex[length=3mm]}] (Modulator) edge node [above, pos=0.40] {$\tilde{\boldsymbol{X}}$} (Channel);
		\path[-{Latex[length=3mm]}] (Modulator) edge node [below, pos=0.40] {\begin{tabular}{c}$\boldsymbol{X}_1,\boldsymbol{X}_2,\dotsc,\boldsymbol{X}_{n}$\\ Length $Mn$ vector \end{tabular}} (Channel);
		\path[-{Latex[length=3mm]}] (Modulator) edge node [above,very near end] {$X_i$} (Channel);
		\path[-{Latex[length=3mm]}] (Channel) edge node [above, very near start]{$Y_i$} (Output);
		\path[-{Latex[length=3mm]}] (Channel) edge node [above, pos=0.65] {$\tilde{\boldsymbol{Y}}$} (Output);
		\path[-{Latex[length=3mm]}] (Channel) edge node [below, pos=0.65] {\begin{tabular}{c}$\boldsymbol{Y}_1,\boldsymbol{Y}_2,\dotsc,\boldsymbol{Y}_{n}$\\ Length $Mn$ vector\\ in $\mathbb{N}_0^{Mn}$\end{tabular}} (Output);		
		\draw[red,thick,dotted] ($(Channel.north west)+(-0.9,0.6)$)  rectangle ($(Channel.south east)+(0.9,-0.6)$);
		\node (tb) [below of=Channel] {\textcolor{red}{Used $Mn$ times serially}};
		\node at (34.5em,0) (aa) {};
		\end{tikzpicture}
	\end{adjustbox}
	\caption{Communication scheme.}
	\label{fig:TScheme}
\end{figure*}
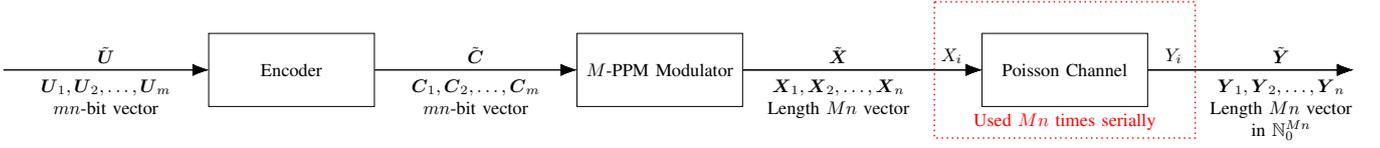
The transmission scheme is depicted in ~\fig{fig:TScheme}. The channel input is a vector $\tilde{\boldsymbol{X}}$ of length $Mn$ where $M$ is the PPM order and $n$ is the number of PPM symbols in one block. The vector $\tilde{\boldsymbol{Y}}$ is the channel output. We model the channel as memoryless, and we have
\begin{equation}
	P_{\tilde{\boldsymbol{Y}}|\tilde{\boldsymbol{X}}}\left(\tilde{\boldsymbol{y}}|\tilde{\boldsymbol{x}}\right)=\prod_{q=1}^{Mn}P_{Y|X}\left(y_q|x_q\right).
\end{equation}  
The channel $P_{Y|X}\left(y|x\right)$ is modeled as a slotted binary input, discrete output Poisson channel. In a given time slot, there is either a pulse, represented by $x=1$, or there is no pulse, represented by $x=0$. Let $n_\text{s}$ be the average number of received photons in a pulsed time slot (when no noise is present) and let $n_\text{b}$ be the average number of received noise photons per slot. Then the conditional probabilities to receive $y \in \mathbb{N}_0=\lbrace 0,1,2,\dotsc\rbrace$ photons follow the Poisson distributions \cite{Barsoum10}:
\begin{align}
\label{eq:unpulsedCH}
P_{Y|X}(y|0)&=\frac{e^{-n_{\text{b}}}n_{\text{b}}^y}{y!}\\
\label{eq:pulsedCH}
 P_{Y|X}(y|1)&=\frac{e^{-(n_{\text{b}}+n_{\text{s}})}(n_{\text{b}}+n_{\text{s}})^y}{y!}.
\end{align}
%
%

\subsection{Pulse Position Modulation}
\label{sec:PPMMod}
PPM is a binary slotted modulation scheme where $M$ is the number of slots \cite{Hemmati05}. To obtain a PPM symbol, $m=\log_2 M$ bits are input to the PPM modulator. Each symbol consists of $M-1$ slots without energy (unpulsed slots) and exactly one slot that contains energy (pulsed slot) \cite{Hemmati05}. We model the $M$-PPM modulator as an orthogonal binary code with rate $\frac{m}{M}$. Each code word is of the form $\boldsymbol{x}^d=\left(0,0,\dotsc,1,\dotsc,0,0\right)$ where the $1$ is in the $d$th position, i.e., $x_d=1$, $x_p=0$, for $p \neq d$ and $p,d\in\lbrace1,2,\dotsc,M\rbrace$. Therefore the modulation alphabet is $\mathcal{X}=\lbrace \boldsymbol{x}^1,\dotsc,\boldsymbol{x}^M\rbrace$, with cardinality $\lvert\mathcal{X}\rvert=M$. The output alphabet is $\mathcal{Y}=\mathbb{N}_0^M$ with an infinite number of elements.

\subsection{Achievable Rates}
The capacity of PPM modulation on the Poisson channel is:
\begin{equation}
\label{eq:cap}
	C^{\text{PPM}}=\frac{1}{M}\text{I}\left(\boldsymbol{X};\boldsymbol{Y}\right)\quad [\text{bits/slot}]
\end{equation}
where $\text{I}\left(\boldsymbol{X};\boldsymbol{Y}\right)$ is the mutual information of the input and the output of the channel, the capacity-achieving input distribution is uniform, i.e., ${P_{\boldsymbol{X}}\left(\boldsymbol{x}\right)=\frac{1}{M}}\:\forall \boldsymbol{x}\in \mathcal{X}$ \cite{Hemmati05}, and $P_{Y|X}$ is defined as in \eref{eq:unpulsedCH} and \eref{eq:pulsedCH}.
The capacity can be expressed as \cite{Barsoum10}:
\begin{equation}
\label{eq:capacity}
C^{\text{PPM}}=\frac{1}{M}\text{E}\left[\log_2\frac{ML\left(Y_1\right)}{\sum_{p=1}^{M}L\left(Y_p\right)}\right]\quad [\textnormal{bits/slot}]
\end{equation}
where $L(y)=\frac{P_{Y|X}(y|0)}{P_{Y|X}(y|1)}$ is the likelihood ratio of the received value $y$, $Y_1$ is distributed as $P_{Y|X}(\cdot|1)$ whereas $Y_p$, for $p\neq 1$, is distributed as $P_{Y|X}(\cdot|0)$. For $n_{\text{b}}>0$ the capacity is:

\begin{align}
\label{eq:cap_nb_gr0}
\begin{split}
C^{\text{PPM}}=&\frac{\log_2\left(M\right)}{M}\\
&-\frac{1}{M}\text{E}\left[\log_2\left(\sum_{p=1}^{M}\left(1+\frac{n_{\text{s}}}{n_{\text{b}}}
\right)^{Y_p-Y_1}\right)\right]\\ 
&[\textnormal{bits/slot}].
\end{split}
\end{align}
For $n_{\text{b}} \to 0$ the expression reduces to
\begin{align}
\label{eq:cap_nb_eq0}
C^{\text{PPM}} = \frac{\log_2\left(M\right)}{M}\left(1-e^{-n_{\text{s}}}\right)\quad [\textnormal{bits/slot}].
\end{align}

\subsection{Polar Coding}
A binary polar code of block length $n$ and dimension $k$ is defined by $n-k$ frozen bit positions and the polar transform $\mathbb{F}^{\otimes \log_2 n}$, which denotes the $\log_2 n$-fold Kronecker power of
\begin{equation}
\mathbb{F}=	
\begin{bmatrix}
1 & 0\\
1 & 1	
\end{bmatrix}.	
\end{equation}
Polar encoding can be represented by 
\begin{equation}
	\boldsymbol{u}\mathbb{F}^{\otimes \log_2 n}=\boldsymbol{c}
\end{equation}
where the $n-k$ frozen positions in $\boldsymbol{u}$ are set to predetermined values and where the unfrozen positions contain $k$ information bits. The vector $\boldsymbol{c}$ is the code word \cite{Arikan09}. Successive cancellation (SC) decoding detects the bits $u_1u_2\cdots u_n$ successively, i.e., the channel output $\boldsymbol{y}$ and the decisions $\hat{u}_1\cdots\hat{u}_i$ are used to detect bit $u_{i+1}$.
\subsection{PPM Mapping and Demapping}

\begin{figure}[!t]
	\centering
	\includegraphics[width=\columnwidth]{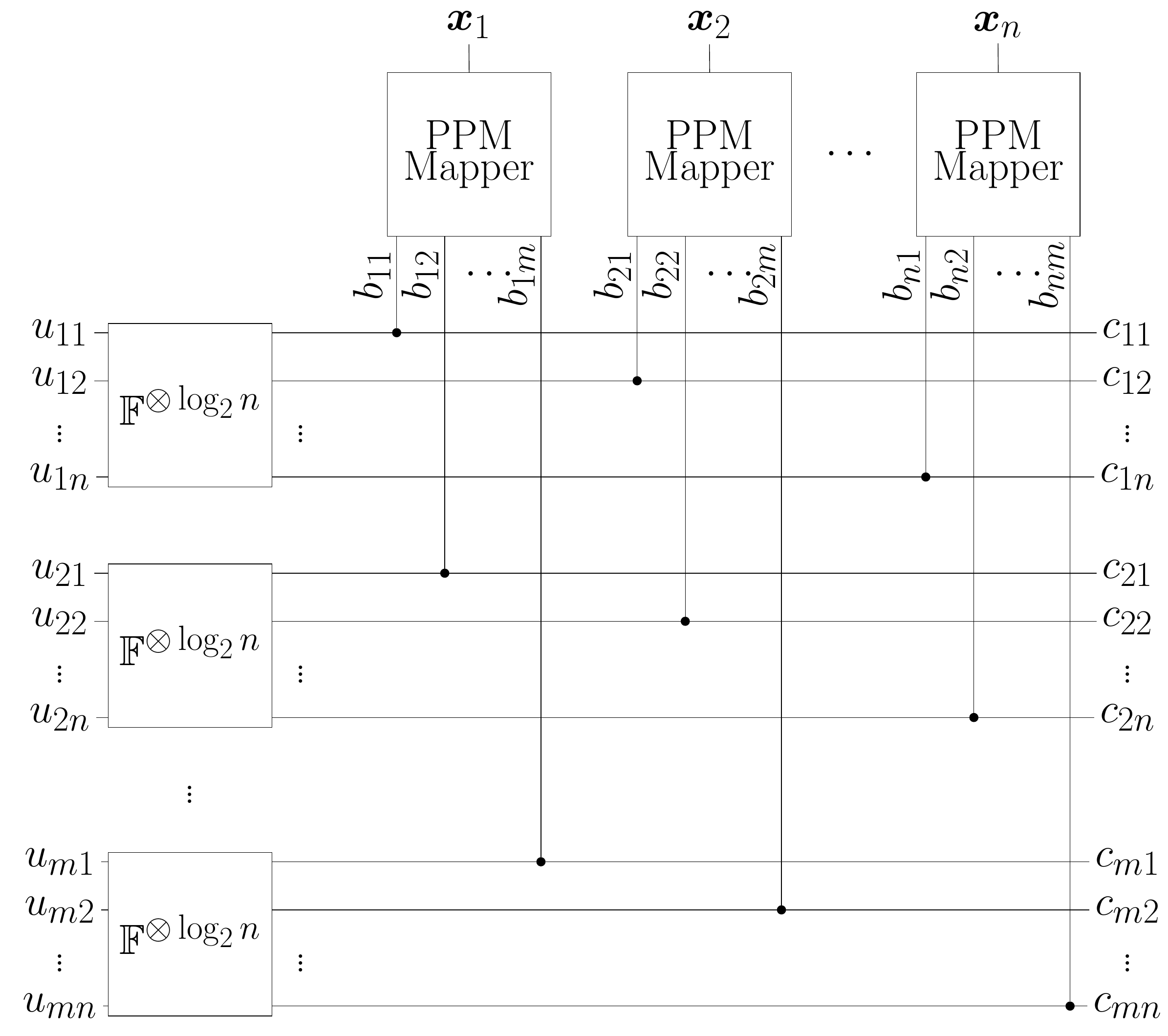}
	\caption{Canonical PCM \cite{Seidl13,Georg16}.}
	\label{fig:CodedMod}
\end{figure}
~\fig{fig:CodedMod} depicts the canonical polar-coded modulation (PCM) \cite{Seidl13,Georg16} instantiated for PPM. The encoder input is a vector $\tilde{\boldsymbol{u}}$ with length $mn$ that contains the information bits and the frozen bits. The vector $\tilde{\boldsymbol{u}}$ is split into $m$ vectors $\boldsymbol{u}_1,\dotsc,\boldsymbol{u}_m$ which are mapped to $\boldsymbol{c}_i=\boldsymbol{u}_i \mathbb{F}^{\otimes \log_2 n}$. The encoding produces the code word $\tilde{\boldsymbol{c}}=\boldsymbol{c}_1,\dotsc,\boldsymbol{c}_m$.

The PPM mapper implements a label function that maps the $m$ bits $c_{1i}\cdots c_{mi}$ to the $i$th transmitted PPM symbol $\boldsymbol{x}_i$ for $i\in \lbrace 1,\dotsc,n\rbrace$, i.e., for $j\in \lbrace 1,\dotsc,m\rbrace$, the output $\boldsymbol{c}_j$ of the $j$th polar transformation is mapped to the $j$th bit level of the input of the labeling function. We define the PPM label as:
\begin{equation}
\boldsymbol{b}_i=b_{i1}\cdots b_{im}:= c_{1i}\cdots c_{mi}, \quad i\in \lbrace 1,\dotsc,n\rbrace.
\end{equation}
To refer to a label at a generic time instance we drop the subscript $i$ and write $\boldsymbol{b}=b_1\cdots b_m$ \cite{Georg16}.

The input $\boldsymbol{x}_i$ to the PPM Poisson channel is the output of the PPM mapper defined as
\begin{align}
\label{eq:func}
f\colon \lbrace0,1 \rbrace^m &\to \mathcal{X}=\lbrace0,1 \rbrace^M\\
\boldsymbol{b} &\to \boldsymbol{x}^{d\left(\boldsymbol{b}\right)}
\end{align}
where 
\begin{equation}
	d\left(\boldsymbol{b}\right)=1+\sum_{j=1}^{m}b_j2^{j-1}
\end{equation}
e.g., for $m=3$, $\boldsymbol{x}^{d\left(010\right)}=00100000$. For a channel code with block length $mn$, $n$ $M$-PPM symbols are transmitted.

\begin{figure}[!t]
	\centering
	\includegraphics[width=\columnwidth]{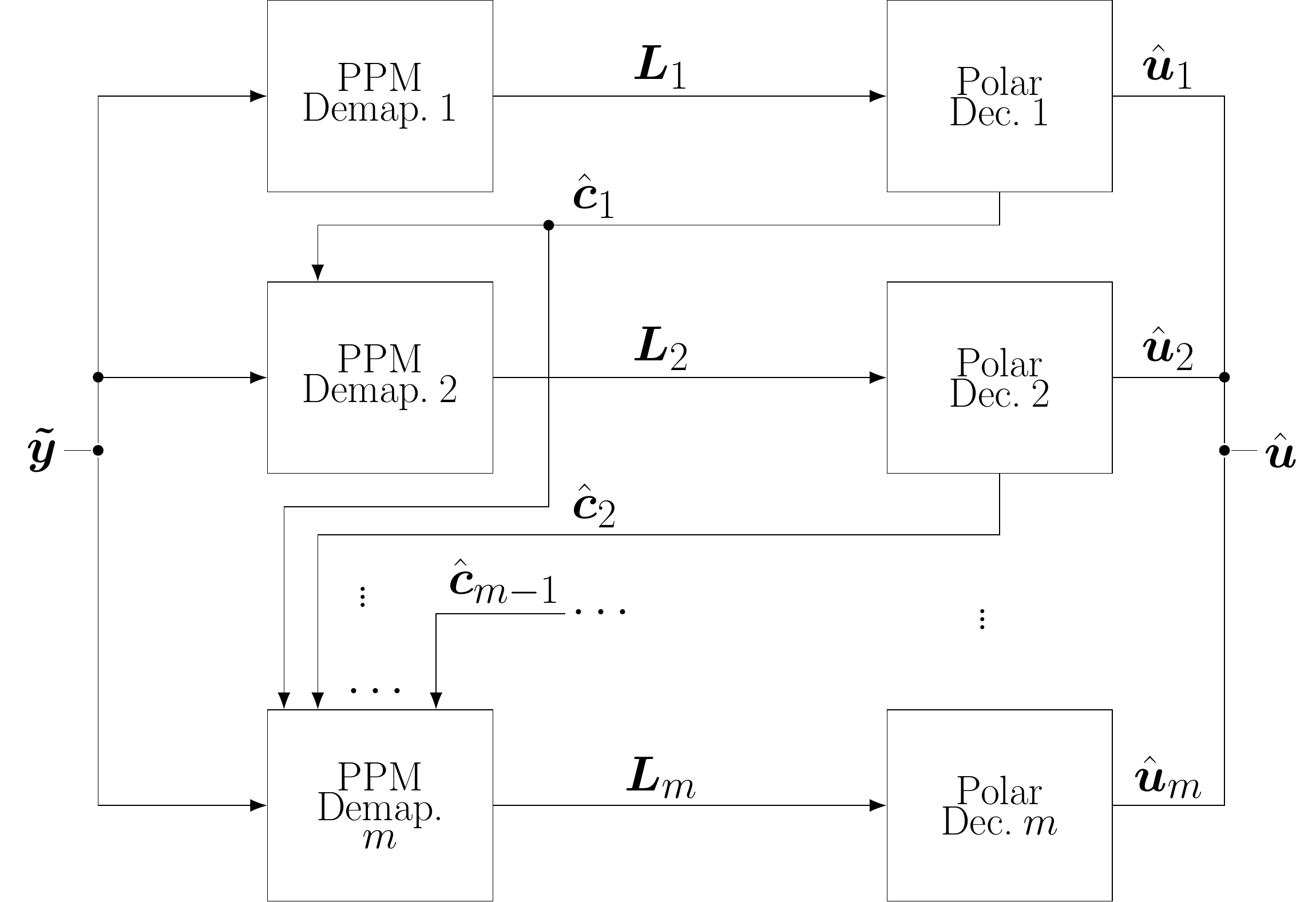}
	\caption{Receiver side of the PCM scheme.}
	\label{fig:MSD}
\end{figure}

The demapping and decoding procedure is depicted in Fig.~\ref{fig:MSD}. The decoding schedule is as follows. First demap the first bit-level, then decode it. Next demap the second bit-level, then decode it. At the $j$th step demap the $j$th bit-level, then decode it. Continue until all $m$ bit-levels have been demapped and decoded. The $j$th PPM demapper uses the channel output $\tilde{\boldsymbol{Y}}$ and the previously detected bit-levels $\hat{\boldsymbol{C}}^{j-1}=\hat{\boldsymbol{C}}_{1},\dots,\hat{\boldsymbol{C}}_{j-1}$ to calculate a soft-information 
\begin{equation}
	\boldsymbol{L}_j=\boldsymbol{\lambda}_j \left(\tilde{\boldsymbol{Y}},\hat{\boldsymbol{C}}^{j-1}\right)
\end{equation}
for the $j$th polar decoder which, in turn, produces a decision $\hat{\boldsymbol{C}}_j$. 
In particular, the calculation for each received symbol $\boldsymbol{y}_i$ at the $j$th level is
\begin{equation}
\begin{split}
\label{eq:LLRl1}
L_{ji}&=\lambda_j\left(\boldsymbol{Y}_i,\hat{\boldsymbol{B}}_{i}^{j-1}\right)\\
&=\log\frac{P_{\boldsymbol{Y}|B_j\boldsymbol{B}^{j-1}}\left(\boldsymbol{Y}_i|B_{ij}=0\hat{\boldsymbol{B}}_{i}^{j-1}\right)}{P_{\boldsymbol{Y}|B_j\boldsymbol{B}^{j-1}}\left(\boldsymbol{Y}_i|B_{ij}=1\hat{\boldsymbol{B}}_{i}^{j-1}\right)}
\end{split}
\end{equation}
where $\hat{\boldsymbol{B}}_{i}^{j-1}=\hat{B}_{i1}\cdots\hat{B}_{i\left(j-1\right)}$.

\begin{example}
Consider the scenario with $M=4$ and the PPM mapping ${\boldsymbol{x}^1=f(00)}$, ${\boldsymbol{x}^2=f(10)}$, ${\boldsymbol{x}^3=f(01)}$, ${\boldsymbol{x}^4=f(11)}$. We calculate $L_{1}\left(\boldsymbol{y}\right)$ at a generic time instant where $\boldsymbol{y}=(y_1,y_2,y_3,y_4)$. By using Bayes' rule and the equiprobable distribution of all label bits, $L_{1}\left(\boldsymbol{y}\right)$ is calculated as
\begin{equation}
	L_{1}\left(\boldsymbol{y}\right)=\log\frac{P_{\boldsymbol{Y}|B_1}\left(\boldsymbol{y}|0\right)}{P_{\boldsymbol{Y}|B_1}\left(\boldsymbol{y}|1\right)}
\end{equation}
where
\begin{equation}
\begin{split}
	P_{\boldsymbol{Y}|B_1}\left(\boldsymbol{y}|0\right)&=\frac{e^{-n_{\text{b}}}n_{\text{b}}^{y_2}}{y_2!}\frac{e^{-n_{\text{b}}}n_{\text{b}}^{y_4}}{y_4!}\\
	&\times\Bigg(\frac{1}{2}\left(\frac{e^{-n_{\text{b}}}n_{\text{b}}^{y_1}}{y_1!}\frac{e^{-\left(n_{\text{s}}+n_{\text{b}}\right)}\left(n_{\text{s}}+n_{\text{b}}\right)^{y_3}}{y_3!}\right)\\
	&+\frac{1}{2}\left(\frac{e^{-\left(n_{\text{s}}+n_{\text{b}}\right)}\left(n_{\text{s}}+n_{\text{b}}\right)^{y_1}}{y_1!}\frac{e^{-n_{\text{b}}}n_{\text{b}}^{y_3}}{y_3!}\right)\Bigg)
\end{split}
\end{equation}

\begin{equation}
\begin{split}
P_{\boldsymbol{Y}|B_1}\left(\boldsymbol{y}|1\right)&=\frac{e^{-n_{\text{b}}}n_{\text{b}}^{y_1}}{y_1!}\frac{e^{-n_{\text{b}}}n_{\text{b}}^{y_3}}{y_3!}\\
&\times\Bigg(\frac{1}{2}\left(\frac{e^{-n_{\text{b}}}n_{\text{b}}^{y_2}}{y_2!}\frac{e^{-\left(n_{\text{s}}+n_{\text{b}}\right)}\left(n_{\text{s}}+n_{\text{b}}\right)^{y_4}}{y_4!}\right)\\
&+\frac{1}{2}\left(\frac{e^{-\left(n_{\text{s}}+n_{\text{b}}\right)}\left(n_{\text{s}}+n_{\text{b}}\right)^{y_2}}{y_2!}\frac{e^{-n_{\text{b}}}n_{\text{b}}^{y_4}}{y_4!}\right)\Bigg).
\end{split}
\end{equation}
Similarly, for $L_2\left(\boldsymbol{y}\right)$:
\begin{equation}
	L_{2}\left(\boldsymbol{y}\right)=\log\frac{P_{\boldsymbol{Y}|B_2B_1}\left(\boldsymbol{y}|0\hat{b}_1\right)}{P_{\boldsymbol{Y}|B_2B_1}\left(\boldsymbol{y}|1\hat{b}_1\right)}.
\end{equation}
Suppose $\hat{b}_1=0$. Then we have
\begin{equation}
\begin{split}
P_{\boldsymbol{Y}|B_2B_1}\left(\boldsymbol{y}|00\right)=&\frac{e^{-n_{\text{b}}}n_{\text{b}}^{y_2}}{y_2!}\frac{e^{-n_{\text{b}}}n_{\text{b}}^{y_3}}{y_3!}\frac{e^{-n_{\text{b}}}n_{\text{b}}^{y_4}}{y_4!}\\
&\times\frac{e^{-\left(n_{\text{s}}+n_{\text{b}}\right)}\left(n_{\text{s}}+n_{\text{b}}\right)^{y_1}}{y_1!}
\end{split}
\end{equation}
\begin{equation}
\begin{split}
P_{\boldsymbol{Y}|B_2B_1}\left(\boldsymbol{y}|10\right)=&\frac{e^{-n_{\text{b}}}n_{\text{b}}^{y_1}}{y_1!}\frac{e^{-n_{\text{b}}}n_{\text{b}}^{y_3}}{y_3!}\frac{e^{-n_{\text{b}}}n_{\text{b}}^{y_4}}{y_4!}\\
&\times\frac{e^{-\left(n_{\text{s}}+n_{\text{b}}\right)}\left(n_{\text{s}}+n_{\text{b}}\right)^{y_2}}{y_2!}.
\end{split}
\end{equation}
\end{example}


\subsection{Successive Cancellation List Decoding}
We use successive cancellation list decoding (SCL) with list size $L \in \mathbb{N}$ \cite{Tal15}. To improve performance, we use an outer CRC code for the information bits. We found experimentally that the 14-CRC code with polynomial $0x27cf$ and the 16-CRC code with polynomial $0xd175$ are good choices \cite{Koopman}. The PPM demappers calculate  soft information for each active path from the previous level.
At the end of the decoding process, the most likely path, among the $L$ paths, that passes the CRC is selected as the decoder's decision. The complexity of the decoder is $O(Lmn\log n)$ \cite{Tal15}.

\section{Polar Code Construction}
The polar code construction encompasses first choosing a desired block length $n$ and a desired rate $R=\frac{k}{n}$ for the code and then finding the set of frozen bits \cite{Arikan09}. There are several ways to construct the codes:
\begin{itemize}
	\item By Monte Carlo simulation.
	\item By using the Binary Erasure Channel (BEC) as a surrogate channel.
	\item By Gaussian approximation.
\end{itemize}
The construction of a polar code via Monte Carlo (MC) simulations is described in \cite{Arikan09}. The MC construction method relies on extensive simulations in order to find the best bit-channels. Polar codes can also be constructed by using the BEC as a surrogate channel, i.e. by replacing the Poisson channel by a BEC channel with the same capacity. Then, by ordering the capacities for each bit-channel the "good" channels are found \cite{Arikan09}. In \cite{Trifonov12} it is shown that the construction of polar codes can be done efficiently by using the Gaussian approximation (GA) construction method.

\subsection{Construction of Polar Codes via biAWGN Surrogate Channel}
\subsubsection{biAWGN Surrogate Channel \cite{Georg16}}
\label{subsec:biAWGN}
The channel is defined as
\begin{equation}
\label{eq:biawgn}
Y = x_b+\sigma Z
\end{equation}
where $x_0 = 1$ and $x_1=-1$ and $Z$ is zero mean Gaussian noise. We define
\begin{equation}
	\text{R}_{\text{biAWGN}}\left(\sigma^2\right)=\text{I}\left(B;x_B+\sigma Z\right)
\end{equation}
where $B$ is uniformly distributed on $\lbrace0,1\rbrace$.
\subsubsection{Gaussian Approximation (GA)}

\begin{figure}[!t]
	\centering
	\includegraphics[width=\columnwidth]{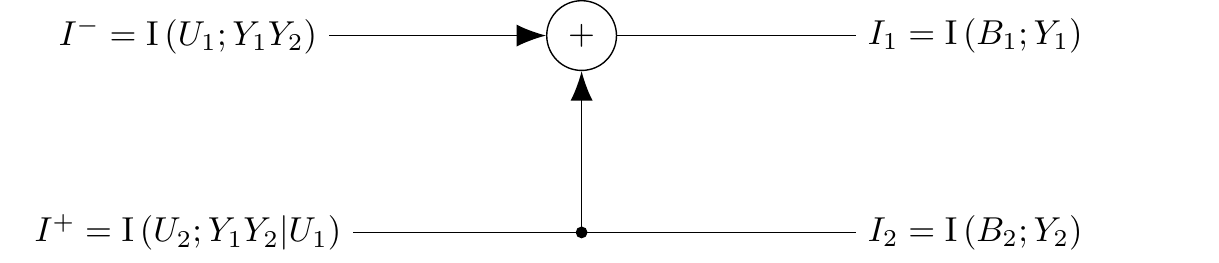}
	\caption{MIs of the basic polar transform \cite{Georg16}.}
	\label{fig:MItrack}
\end{figure}

The GA construction method for polar codes was proposed in \cite{Trifonov12}. The reliability of the bit $U_i,\quad i\in \lbrace0,\dotsc,n\rbrace$, can be quantified by the mutual information $\text{I}\left(U_i;Y^n\given U_i^{i-1}\right)$. We can calculate these MIs for all $i \in \lbrace1,\dotsc,n\rbrace$ by recursively calculating the MIs of the basic polar transform shown in ~\fig{fig:MItrack}. For the biAWGN channel the update rule for the basic polar transform is \cite{Kramer04}

\begin{equation}
I^-=1-J\left(\sqrt{\left[J^{-1}\left(1-I_1\right)\right]^2+\left[J^{-1}\left(1-I_2\right)\right]^2}\right)
\end{equation}

\begin{equation}
I^+=J\left(\sqrt{\left[J^{-1}\left(I_1\right)\right]^2+\left[J^{-1}\left(I_2\right)\right]^2}\right)
\end{equation}
where the $J$ function is
\begin{equation}
J\left(\sigma\right)=1-\int_{-\infty}^{\infty}\frac{e^{-\frac{\left(\xi-\frac{\sigma^2}{2}\right)^2}{2\sigma^2}}}{\sqrt{2\pi}\sigma}\log_2\left(1+e^{-\xi}\right)\mathrm{d}\xi.
\end{equation}
To calculate the $J$ function and its inverse $J^{-1}$, we use the approximation
\begin{equation}
J\left(\sigma\right)\approx \left(1-2^{-H_1\sigma^{2H_2}}\right)^{H_3}
\end{equation}
\begin{equation}
J^{-1}\left(I\right)\approx\left(-\frac{1}{H_1}\log_2\left(1-I^{\frac{1}{H_3}}\right)\right)^{\frac{1}{2H_2}}
\end{equation}
from \cite[Eqs. (9), (10)]{Brannstrom05} where $H_1 = 0.3073$, $H_2=0.8935$ and $H_3=1.1064$.
\subsubsection{MI Demapper Gaussian Approximation Construction}
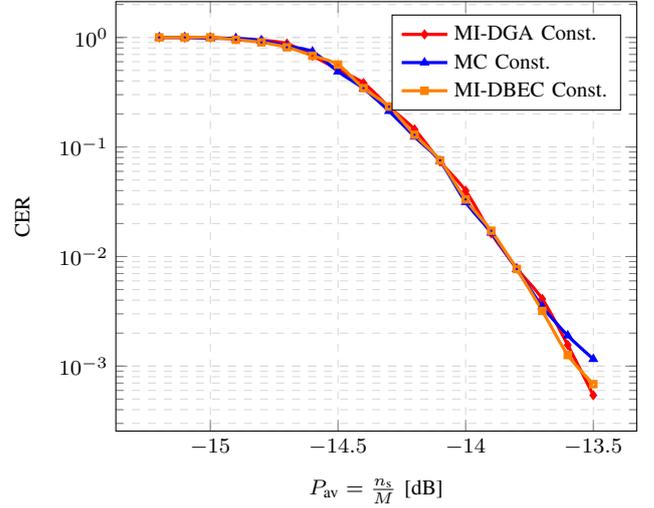
\begin{figure}[!t]
	\centering
	\footnotesize
	\begin{tikzpicture}
	\begin{semilogyaxis}[
	legend pos=  north east,
	xlabel={$P_\text{av}=\frac{n_\text{s}}{M}$ [dB]},
	ylabel={CER},
	grid=both,
	legend cell align=left,
	grid style={dashed, gray!30},
	every axis plot/.append style={very thick}
	]
	\addplot[red] plot[smooth, mark=diamond, mark size=1pt]  table [x={P}, y={Pe}] {GA_FER1.dat};
	\addlegendentry{MI-DGA Const.}	
	\addplot[blue] plot[smooth ,mark=triangle, mark size=1pt] table [x={P}, y={Pe}] {MC_FER.dat}; 
	\addlegendentry{MC Const.}
	\addplot[orange] plot[smooth, mark=square, mark size=1pt] table [x={P}, y={Pe}] {BEC_FER.dat}; 
	\addlegendentry{MI-DBEC Const.}
	\end{semilogyaxis}
	\end{tikzpicture}
	\caption{Comparison of MC, MI-DGA and MI-BEC construction methods. The block length is $6144$, the code rate is $1/2$ and $n_\text{b}=0.2$.}
	\label{fig:MCconstComp}
\end{figure}
For the proposed scheme, we use the construction approach from \cite{Georg16} called MI demapper GA (MI-DGA). The idea is to first characterize the bit-channels at the output of the PPM demappers by mutual information expressions that take into account the dependence of the soft information produced by the PPM demappers on all the previous detected bits.  Then we replace the PPM demapper bit-channels by biAWGN surrogate channels with the corresponding MI and use the GA construction with the $J$ function defined above. 

Using the chain rule for mutual information we have
	\begin{align}
\label{eqn:smrate}
\begin{split}
\text{I}(\boldsymbol{X};\boldsymbol{Y})=\text{I}(\boldsymbol{B};\boldsymbol{Y})&=\sum_{j=1}^{m}\text{I}(B_j;\boldsymbol{Y}|\boldsymbol{B}^{j-1})\\
&=\text{I}(B_1;\boldsymbol{Y})+\text{I}(B_2;\boldsymbol{Y}|B_1)+\ldots\\
&+\text{I}(B_m;\boldsymbol{Y}|B_1\cdots B_{m-1}).
\end{split}
\end{align}
Observe that the $j$th bit-level has the MI $\text{I}(B_j;\boldsymbol{Y}|B_1\cdots B_{j-1})$. Therefore, the MI-DGA costruction calculates $\text{I}(B_j;\boldsymbol{Y}|B_1\cdots B_{i-1})$ for all $j \in \lbrace1,2,\dots,m\rbrace$ and connects $L_j$ with $B_j$ by a biAWGN channel with noise variance \cite{Georg16}
\begin{equation}
	\sigma_j^2\colon \text{R}_{\text{biAWGN}}\left(\sigma_j^2\right)=\text{I}\left(B_j;\boldsymbol{Y}|\boldsymbol{B}^{j-1}\right).
\end{equation}
Then, we use the GA to find the most reliable bits in $\tilde{\boldsymbol{u}}$. To construct a length $mn$ and rate $R=k/n$ polar code, find the set of $nm-km$ most unreliable bits in $\tilde{\boldsymbol{u}}$ and freeze them.

\begin{remark}
Observe that in \eref{eqn:smrate} on each bit-level the conditional MI is an achievable rate \cite{Gallager} and the sum-rate of all the bit-levels is exactly the capacity of the PPM. Therefore $\text{C}^{\text{PPM}}$ is an achievable rate with our scheme. Note that if in \eref{eqn:smrate} we disregard the conditioning on the previous bits of the symbol, then we calculate the BMD rate \cite[Eq. (10)]{Martinez}
	\begin{equation}
	\label{eqn:bmrate}
	\sum_{j=1}^{m}\text{I}(B_j;\boldsymbol{Y}|\boldsymbol{B}^{j-1})\geq\sum_{j=1}^{m}\text{I}(B_j;\boldsymbol{Y})=R^{\text{BMD}.}
	\end{equation}
 ~\fig{fig:cm_vs_bicm1} plots $R^{\text{BMD}}$ vs. $P_{\text{av}}$ .
\end{remark}

\subsection{Construction of Polar Codes for the PPM Poisson Channel via the BEC Surrogate Channel}

We introduce a MI demapper BEC (MI-DBEC) construction for polar codes on the PPM Poisson channel. We follow the same idea as for the MI-DGA. Therefore, the MI-DBEC construction calculates $\text{I}(B_j;\boldsymbol{Y}|B_1\cdots B_{i-1})$ for all $j \in \lbrace1,2,\dots,m\rbrace$ and connects $L_j$ with $B_j$ by a BEC channel with erasure probability
\begin{equation}
\epsilon_j=1-\text{I}\left(B_j;\boldsymbol{Y}|\boldsymbol{B}^{j-1}\right).
\end{equation}
We then use the construction for the BEC from \cite{Arikan09}.
 
\subsection{Comparison of Polar Code Constructions}

To verify that the results in \cite{Georg16} extend to the PPM Poisson channel, we compare the performance of a polar code constructed via the MC approach (this code is always going to be \enquote{good}) with the performance of a polar code obtained via MI-DGA and MI-DBEC. ~\fig{fig:MCconstComp} plots the codeword error rate (CER) performance curves for a polar code designed with MC simulations, for a polar code designed via MI-DBEC, and for a polar code designed with MI-DGA. The block length is $n=6144$ and the rate is $R=1/2$. The average number of noise photons per slot is $n_\text{b}=0.2$. The MC, MI-DGA and MI-DBEC curves virtually coincide. We use the MI-DGA construction to design the codes. However, the MI-DBEC construction could also be used.

\section{Numerical Results}
\label{sec:results}
\begin{table*}[!t]
	\centering
	\caption{Distribution of the List Size Used in SC List Decoding}
	\label{table:1}
	\begin{tabular}{||c|| c | c | c | c | c | c||} 
		\hline
		\backslashbox{$L$}{$\frac{n_{\text{s}}}{M}$ [dB]}& $-15.2$ & $-15.1$ & $-15.0$ & $-14.9$ & $-14.8$ & $-14.7$ \\ 
		\hline\hline
		32       & 12  & 53  & 292& 2426& 11454& 149180\\ 
		\hline
		64       & 2  & 22  & 66  & 294 & 670    & 3144\\ 
		\hline
		128     &  4 & 23  & 56  & 209  & 411    & 1719\\ 
		\hline 
		256     & 5  & 17  & 51  & 181  & 239    & 910\\ 
		\hline
		512     & 7  & 17  & 37  & 117   & 157    & 545\\ 
		\hline
		1024   & 6  & 16  & 44  & 70    & 89   & 304\\ 
		\hline
		2048   & 4  & 15  & 15  & 58  & 61    & 175\\ 
		\hline
		4096   & 4  & 7    & 17  & 31    & 40      & 114\\ 
		\hline
		8192   & 4  & 13    & 8  & 28    & 25      & 64\\ 
		\hline
		16384 & 56 & 62 & 58  & 68    & 66     & 81\\ 
		\hline
	\end{tabular}
\end{table*}

\begin{figure}[!t]
	\centering
	\footnotesize
	\begin{tikzpicture}
	\begin{semilogyaxis}[
	legend pos=  north west,
	xlabel={$P_\text{av}=\frac{n_\text{s}}{M}$ [dB]},
	ylabel={BER},legend style={font=\footnotesize},
	grid=both,
	legend cell align=left,
	grid style={dashed, gray!30},
	every axis plot/.append style={very thick}
	]
	\addplot[black] plot[smooth] table [x={P}, y={Pe}] {cap.dat}; 
	\addlegendentry{capacity}
	\addplot[red] plot[smooth ,mark=triangle, mark size=1pt] table [x={P}, y={Pe}] {scppm1.dat}; 
	\addlegendentry{SCPPM}
	\addplot[blue] plot[smooth ,mark=square, mark size=1pt] table [x={P}, y={Pe}] {pc_ls16000.dat}; 
	\addlegendentry{PCM}
	\end{semilogyaxis}
	\end{tikzpicture}
	\caption{Comparison of the proposed PCM scheme with the SCPPM scheme in \cite{Barsoum10}. Both codes have a block length of $nm=8208$ bits and the rate is $R=1/2$. The maximal allowed list size for the PCM code is $L=16384$ and 14-CRC is used.}
	\label{fig:results}
\end{figure}
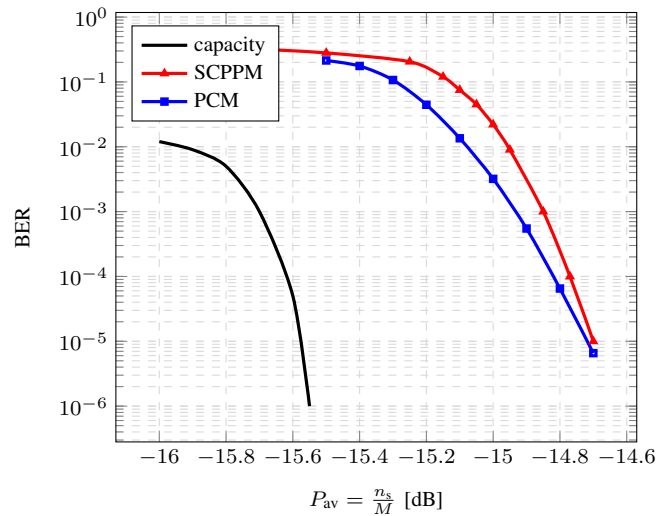

\begin{figure}[!t]
	\centering
	\footnotesize
	\begin{tikzpicture}
	\begin{semilogyaxis}[
	legend pos=  south west,
	xlabel={$P_\text{av}=\frac{n_\text{s}}{M}$ [dB]},
	ylabel={CER},legend style={font=\footnotesize},
	grid=both,
	legend cell align=left,
	grid style={dashed, gray!30},
	every axis plot/.append style={very thick}
	]
	\addplot[red] plot[smooth ,mark=diamond, mark size=1pt] table [x={P}, y={Pe}] {Balasz_0002.dat}; 
	\addlegendentry{NB-LDPC}
	\addplot[blue] plot[smooth ,mark=triangle, mark size=1pt] table [x={P}, y={Pe}] {GA_0002.dat}; 
	\addlegendentry{$\text{PCM}_\text{MI-DGA}$}
	\addplot[orange] plot[smooth ,mark=square, mark size=1pt] table [x={P}, y={Pe}] {BEC_0002.dat}; 
	\addlegendentry{$\text{PCM}_\text{MI-DBEC}$}
	\addplot[red] plot[smooth ,mark=diamond, mark size=1pt] table [x={P}, y={Pe}] {Balasz_02.dat}; 
	\addplot[blue] plot[smooth ,mark=triangle, mark size=1pt] table [x={P}, y={Pe}] {GA_02.dat}; 
	\addplot[orange] plot[smooth ,mark=square, mark size=1pt] table [x={P}, y={Pe}] {BEC_02.dat}; 
	\addplot[red] plot[smooth ,mark=diamond, mark size=1pt] table [x={P}, y={Pe}] {Balasz_2.dat}; 
	\addplot[blue] plot[smooth ,mark=triangle, mark size=1pt] table [x={P}, y={Pe}] {GA_2.dat}; 
	\addplot[orange] plot[smooth ,mark=square, mark size=1pt] table [x={P}, y={Pe}] {BEC_2.dat}; 
	\node[coordinate] (A) at (axis cs:-18.6,0.3) {};
	\node[coordinate,pin = {45:$n_{\text{b}}=0.002$}] at (axis cs:-18.5,0.38){};
	\node[coordinate] (B) at (axis cs:-15.1,0.3) {};
	\node[coordinate,pin = {45:$n_{\text{b}}=0.2$}] at (axis cs:-15,0.38){};
	\node[coordinate] (C) at (axis cs:-11.6,0.3) {};
	\node[coordinate,pin = {-135:$n_{\text{b}}=2$}] at (axis cs:-11.8,0.24){};
	\end{semilogyaxis}
	\draw[black] (A) ellipse (0.4 and 0.2);
	\draw[black] (B) ellipse (0.4 and 0.2);
	\draw[black] (C) ellipse (0.4 and 0.2);
	\end{tikzpicture}
	\caption{Comparison of the proposed PCM scheme with both MI-DGA and MI-DBEC constructions with the NB-LDPC scheme in \cite{Balazs17}. All codes have a block length of $nm=8208$ bits, the rate is $R=1/2$, $M=64$ and $n_{\text{b}} \in\lbrace0.002,0.2,2\rbrace $. The maximal allowed list size for the PCM code is $L=16384$ and 16-CRC is used.}
	\label{fig:results_Balasz}
\end{figure}
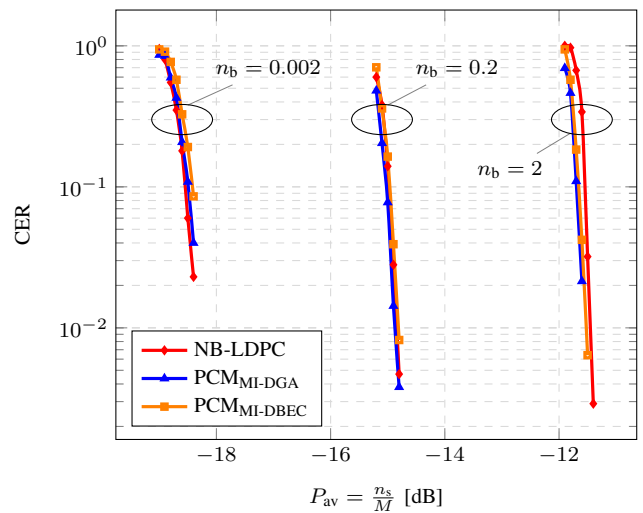

The PCM scheme described above was implemented and the results are compared with the best scheme in \cite{Barsoum10} where the rate loss entailed by BMD is mitigated by introducing iterations between the (outer) decoder and the PPM demapper. By doing so, a BMD-ID scheme is realized where the outer code is a convolutional code serially concatenated with the PPM demapper (which is modified by embedding in it a binary accumulator). This scheme is referred to as SCPPM.

~\fig{fig:results} depicts the simulation results. The SCPPM code proposed in \cite{Barsoum10} has a block length of $8208$ bits and rate $R=1/2$ (orange curve in ~\fig{fig:results}). The PCM scheme proposed in this paper (red curve in ~\fig{fig:results}) has a block length of $nm=8208$ bits, with $k=4104$ information bits. An outer $14$-CRC code is concatenated to the information bits, thus making the polar coding rate $R=(4104+14)/(8208)$, and the overall rate $R=1/2$. A dynamic list decoder is used with a maximal allowed list size of $L=16384$. We define the notion of dynamic list size as follows:
\begin{itemize}
	\item Start decoding with list size $L=32$.
	\item If none of the candidates passes the CRC test, the list size is doubled and the decoding is started once again.
	\item Keep doubling the list size either until a code word that passes the CRC (valid code word) is found or until the limit of $16384$ is exceeded.
\end{itemize}
The performance simulation for both schemes is done with $64$-PPM and the background noise is $n_\text{b}=0.2$. The parameters are chosen as suitable for a Mars-Earth downlink \cite{Barsoum10}. The proposed scheme achieves a better performance than the best code proposed in \cite{Barsoum10}. For the particular parameters, the non-binary LDPC (NB-LDPC) scheme from \cite{Balazs17} does not show any improvement over the SCPPM scheme, therefore we compare our results only with the SCPPM scheme. For all of the considered average powers and for all of the simulations presented in this work, the stopping criterion for the simulation was to collect $50$ erroneous frames. Table \ref{table:1} shows the distribution of the list size needed to find a valid code word. For example, at an average power of ~\SI{-14,9}{dB}, $2426$ code words were decoded using a list of size $L=32$ and $68-50=18$ code words needed a list size of $L=16384$ to be decoded ($50$ code words were decoded erroneously even at the maximal list size). For a low average power, the decoder usually resorts to the maximal list size to find a valid code word and even then very few code words are decoded correctly. However, as the power increases, the decoder can find a valid code word with smaller list sizes and  at an average power of \SI{-14.7}{dB} a large fraction of code words are decoded with a list size of just $32$. However, even though most of the code words are decoded with a list size of $32$, having a bigger maximal available list size than $16384$ will improve the scheme.

~\fig{fig:results_Balasz} plots the CER of the PCM scheme for different choices of $n_{\text{b}}$ and compares the results with the NB-LDPC scheme from \cite{Balazs17}. The NB-LDPC scheme performs really well when $n_{\text{b}}$ is quite small, thus making the Poisson channel behave like an erasure channel (\cite{Balazs17} uses EXIT analysis on the erasure channel to construct the NB-LDPC codes). However, as $n_{\text{b}}$ increases the PCM scheme starts to gain in performance and, in particular, when $n_{\text{b}}=2$ there is about \SI{0.1}{dB} gain in $P_\text{av}$ with respect to the NB-LDPC scheme.

\subsection{The Impact of the List Size}
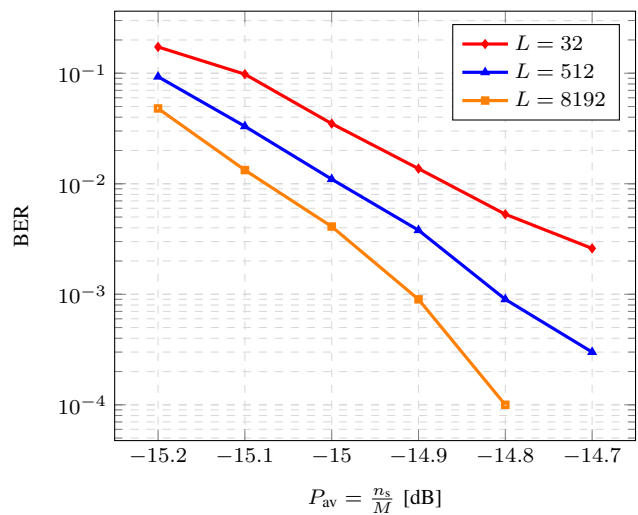
\begin{figure}[!t]
	\centering
	\footnotesize
	\begin{tikzpicture}[]
	\begin{semilogyaxis}[xlabel={$P_\text{av}=\frac{n_\text{s}}{M}$ [dB]}, ylabel={BER},
	legend entries={$L=32$,$L=512$,$L=8192$},
	legend cell align=left,
	legend pos= north east,grid = both,grid style={dashed, gray!30},,every axis plot/.append style={very thick}]
	\addplot [red,mark=diamond, mark size=1pt] table [x={pow}, y={ls32}] {LScompCRC16.dat};
	\addplot [blue,mark=triangle, mark size=1pt] table [x={pow}, y={ls512}] {LScompCRC16.dat};
	\addplot [orange,mark=square, mark size=1pt] table [x={pow}, y={ls8192}] {LScompCRC16.dat};
	\end{semilogyaxis}
	\end{tikzpicture}
	\caption{Dependence between the list size and the BER.}
	\label{fig:lsdep}
\end{figure}
~\fig{fig:lsdep} shows that as the list size increases, the performance of the code becomes better. The codes in ~\fig{fig:lsdep} have a block length of $8028$ bits, a $16$-CRC outer code with polynomial $0x8d95$ \cite{Koopman} and $n_\text{b}=0.2$. Bearing in mind the application of the proposed scheme, the decoding can be done offline, therefore a big list size is not a major problem.

\section{Conclusion}
In this paper, polar coded modulation (PCM) was applied to the PPM Poisson channel. The interplay between the encoder and the modulator was examined and the results in \cite{Georg16} were extended to the PPM Poisson channel. The results show a slight gain in performance with respect to the state of the art transmission scheme proposed in \cite{Barsoum10}. Additionally, we demonstrated that the existing approaches of designing polar codes for the AWGN channel via biAWGN channel surrogates \cite{Georg16} can be extended to design polar codes on the PPM Poisson channel. We observed that the list size of the dynamic successive cancellation list decoder has an impact on the performance of the scheme. Additionally, the polar code construction for list decoding is an interesting direction for further research \cite{Bioglio,Peihong} as it may lead to improved results for the PCM scheme.  
\section*{Acknowledgment}
The authors would like to thank Dr. Gianluigi Liva for his helpful comments.

\bibliographystyle{IEEEtran}
\bibliography{IEEEabrv,literature}
\end{document}